\documentclass[conference,10pt]{IEEEtran}
\usepackage{epsfig,rotating,setspace,latexsym,amsmath,epsf,amssymb,amsfonts,bm,subfigure,epstopdf,cite,authblk,bbm,mathtools,amsthm,algorithm,color,systeme,mathtools,amsmath,derivative,bigints}
\usepackage[
top=1.86 cm,
bottom = 2.75cm,
left   = 1.6cm,
right  = 1.9cm]{geometry}
\usepackage[utf8]{inputenc}
\usepackage[T1]{fontenc}
\usepackage[thinc]{esdiff}
\usepackage[noend]{algpseudocode}
\theoremstyle{plain}
\newtheorem{theorem}{Theorem}
\newtheorem{lemma}{Lemma}
\newtheorem{corollary}{Corollary}

\newtheorem{remark}{Remark}

\algrenewcommand\algorithmicforall{\textbf{foreach}}
\algrenewcommand\algorithmicindent{.8em}
\usepackage{tabularx}

\IEEEoverridecommandlockouts
\allowdisplaybreaks

\begin{document}

\title{Equidistant-Sample or Wait-and-Sample to Minimize Age Under Sampling Constraint?}

\author{Subhankar Banerjee \qquad Sennur Ulukus\\
\normalsize Department of Electrical and Computer Engineering\\
\normalsize University of Maryland, College Park, MD 20742\\
\normalsize \emph{sbanerje@umd.edu} \qquad \emph{ulukus@umd.edu} }

\maketitle

\begin{abstract}
   We study a status update system with a source, a sampler, a transmitter, and a monitor. The source governs a stochastic process that the monitor wants to observe in a timely manner. To achieve this, the sampler samples fresh update packets which the transmitter transmits via an error prone communication channel to the monitor. The transmitter can transmit without any constraint, i.e.,  it can transmit whenever an update packet is available to the transmitter. However, the sampler is imposed with a sampling rate constraint. The goal of the sampler is to devise an optimal policy that satisfies the resource constraint while minimizing the age of the monitor.  We formulate this problem as a constrained Markov decision process (CMDP). We find several structures of an optimal policy. We leverage the optimal structures to find a low complexity optimal policy in an explicit manner, without resorting to complex iterative schemes or techniques that require bounding the age. 
\end{abstract}

\section{Introduction}
\begin{figure}[t]
    \centerline{\includegraphics[width = 0.9\columnwidth]{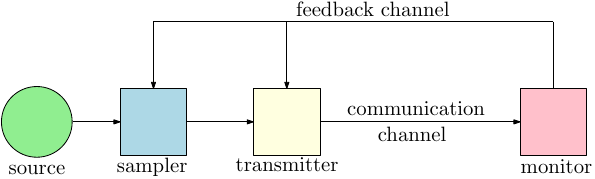}}
    \caption{Different network components for the considered system model.}
    \label{fig:1}
\end{figure}

We consider a network model, where a source governs a stochastic process and a monitor aims to track the source in a timely manner. We consider the well-known metric in the literature, \emph{age of information}, to measure the freshness of the information that the monitor has about the source \cite{YatesSurvey}. The sampler samples a new update packet from the source and passes it to the transmitter. The transmitter transmits the most recently available packet to the monitor via an erroneous channel. We assume that the sampling process is costly, and thus the sampler has to follow a sampling constraint. However, the transmitter does not have any such constraint, thus, it can transmit whenever there is an available packet. Hence, our study applies to a system where the sampling process is costlier than the transmission process. A pictorial representation of the system model is given in Fig.~\ref{fig:1}. 

In the literature, there are works that study a constrained scheduling problem to minimize age \cite{hatami2022demand, tang2020minimizing, wang2019minimizing, ceran2019average, chen2023minimizing}. However, those works consider a constraint on the transmission, which is either on the total number of transmissions or on the total transmission power. In this work, we separate the sampler and the transmitter, as in many applications both tasks may incur different costs. Note that, the works \cite{hatami2022demand, tang2020minimizing, wang2019minimizing, ceran2019average, chen2023minimizing}, consider a slotted communication channel and not a server with a random delay.

We formulate the problem as a CMDP. We find several structures of an optimal policy for the CMDP. Unlike the existing literature regarding age minimization with CMDP \cite{hatami2022demand, tang2020minimizing, wang2019minimizing, ceran2019average, chen2023minimizing, agheli2024effective}, we find an optimal policy without bounding the age and without using an iterative algorithm, which may be complex. We use the structures of an optimal policy to find an optimal policy explicitly. The method and the proof techniques that we develop to find an optimal policy in this work can also be extended and used to find an optimal policy for an age minimization problem with a single constraint.

If there is no sampling constraint, then always sampling would be an optimal policy as the transmitter will always transmit a fresh update packet, and the age of the monitor drops down to $1$ upon every successful transmission. However, with the sampling constraint, a possible intuition is that it may be optimal that the sampler samples more often when the age of the monitor is high, and samples less often when the age of the monitor is comparatively low. Another intuition is under an optimal policy, the sampler samples an update packet equidistantly over the time horizon, independent of the age of the monitor, while maintaining the sampling constraint.

We find that for our problem, an optimal policy follows the second intuition. We show that there exists either an optimal policy, which equidistantly samples over the time horizon while satisfying the constraint with equality, or there exist two equidistant sampling policies, where one policy strictly satisfies the constraint and the other does not satisfy the constraint, and we can choose randomization between them such that the constraint is satisfied with equality and the randomized policy is an optimal policy. We will see that the optimal sampling policy is independent of the channel erasure probability. The sampling instances are also independent of the feedback, thus given a constraint the optimal sampling instances are fixed and independent of everything else. Due to space limitations here, we skip detailed proofs for the results here, which will be provided in a journal version. In the journal version, we also show that if there is no constraint on the sampler while there is a constraint on the transmitter, then the first intuition would be correct, i.e., a wait-and-sample policy would be optimal in that case.  

Note that, we can think of the transmitter and the communication channel together as a server with a geometric service time. Thus, the problem we consider in this paper can be reformulated as an optimal sampling problem with a preemptive server with a geometric service time. Note that, considering a preemptive server is necessary as in our problem whenever the sampler samples a fresh packet, the staler packet at the transmitter gets replaced by the fresh packet. In the literature, optimal sampling problems with a constraint and a non-preemptive service queue have been considered in \cite{sun2019sampling, sun2019sampling2}. The optimal sampling policy for a preemptive server without a constraint has been considered in \cite{arafa2019aller, banerjee2024preempt}. Note that, the erasure communication channel cannot be thought of as a preemptive server, if we consider a problem with a constraint on the transmitter, as we always assume that a preemption only occurs when a fresher packet arrives at the server \cite{banerjee2024preempt, arafa2019aller, wang2019preempt, yates2018age2}. Thus, to the best of our knowledge, this is the first work to consider the optimal sampling problem in a preemptive server with a constraint, where the service time is restricted to the geometric distribution. An interesting future work would be to consider a similar problem with a more general (i.e., not necessarily geometric) service time. Note that, for a given sampling constraint, the analysis of age, under the equidistant sampling policy is similar to the analysis of age with an $D/Geo/1$ queue, where $D$ stands for deterministic arrival of update packets, and $Geo$ stands for geometric service time.   

\section{System Model}
We denote the sampling decision at time slot $t$, with $a(t)$, which takes values from the set $\{0,1\}$. The decision $a(t)=0$ implies that the sampler does not sample a packet at time $t$, where $a(t)=1$ implies that the sampler samples a packet at time $t$. Thus, a sampling policy $\pi$ is completely defined by the following sequence,
\begin{align}
    \{a^{\pi}(1), a^{\pi}(2), a^{\pi}(3), \cdots\}.
\end{align}
We have the following constraint,
\begin{align}
    \liminf_{T\rightarrow\infty}\frac{1}{T}\sum_{t=1}^{T}\mathbb{E}[a^{\pi}(t)] \leq f_{\max}. 
\end{align}

\begin{figure}[t]
    \centerline{\includegraphics[width = 1\columnwidth]{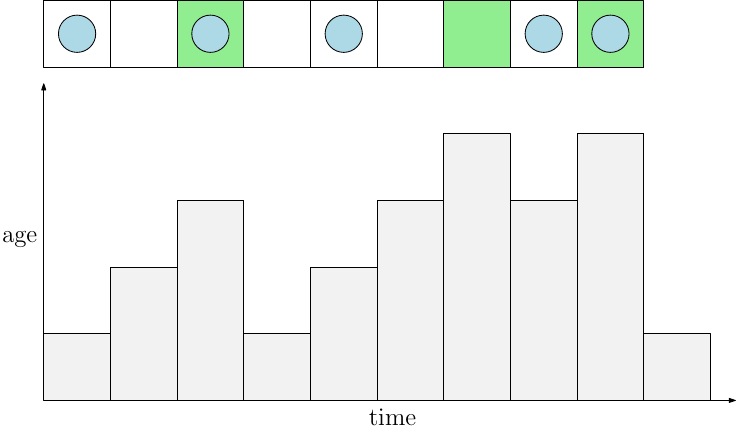}}
    \caption{The cyan circles imply sampling of update packets and green boxes imply successful transmission of the recently sampled update packet.}
    \label{fig:2}
\end{figure}

On the other hand, the transmitter has no constraint, consequently, if there is a packet available at the transmitter, it transmits to the monitor right away. The channel between the transmitter and the monitor is erroneous, where the probability of successful transmission is $q>0$. We assume that a successful packet transmission takes one time slot. Thus, the sampling of a packet occurs at the start of a slot, the transmission of a packet begins at the start of a slot and ends at the end of a slot. Under a sampling policy $\pi$, we denote the age of the monitor at time $t$ with $v^{\pi}(t)$. Whenever a successful transmission occurs, the age of the monitor drops to the generation time of the successfully received packet. We assume that the feedback channel is error-free and instantaneous, i.e., the sampler and the transmitter get to know at the end of a slot if the packet has been successfully delivered to the monitor. A pictorial representation of the age process under an arbitrary sampling policy is given in Fig,~\ref{fig:2}. 

We are interested in the following problem,
\begin{align}\label{eq:3}
    &\inf_{\pi\in\Pi} \limsup_{T\rightarrow \infty} \frac{1}{T} \sum_{t=1}^{T} \mathbb{E}_{\pi} [v^{\pi}(t)|v^{\pi}(1)=v]\nonumber \\  
     & \textrm{s.t.}  \quad   \liminf_{T\rightarrow\infty}\frac{1}{T}\sum_{t=1}^{T}\mathbb{E}[a^{\pi}(t)|v^{\pi}(1)=v] \leq f_{\max}, 
\end{align}
where we assume that $v$ is the initial age of the monitor. 

\section{Optimal Policy}
We reformulate (\ref{eq:3}) in a CMDP framework, then we optimally solve the CMDP. First, we define several components of this CMDP.

\textit{State:} We denote a state $s$ of the system with a vector $s=(v_{1},v_{2})$, where $v_{1}$ corresponds to the age of the monitor and $v_{2}$ corresponds to the age of the packet in service. Note that, the age of packet can never exceed the age of the monitor, i.e., $v_{2}\leq v_{1}$. We denote the collection of all states with state space, $\mathcal{S}$,
\begin{align}\label{eq:4}
    \mathcal{S}=\{(v_{1},v_{2})\ | \ v_{2}\leq v_{1}, (v_{1},v_{2})\in{\mathbb{N}^{2}}\}.
\end{align}

\textit{Action space:} The action space $\mathcal{A}$, contains two vales $a=0$ and $a=1$. Action $a=0$ implies that the sampler does not sample a packet and action $a=1$ implies otherwise. Note that, for both the actions the transmitter transmits the update packet if it has an available update packet. 

\textit{Cost:} For a state $s$ and action $a$, we denote the cost with $C(s,a)$, where $C(s,a)=v_{1}$. Note that, the cost of the system is independent of the action.
Thus, (\ref{eq:3}) can be reformulated as
\begin{align}\label{eq:5} 
    &\inf_{\pi\in{\Pi}}  \limsup_{T\rightarrow \infty} \frac{1}{T} \sum_{t=1}^{T} \mathbb{E}_{\pi} [C(s(t),a(t))|s(1)=(v,\infty)] \nonumber\\ & \textrm{s.t.}  \quad \liminf_{T\rightarrow \infty } \frac{1}{T} \sum_{t=1}^{T} \mathbb{E}_{\pi}[\bm{1}_{a(t)=1}|s(1)=(v,\infty)] \leq f_{\max},
\end{align}
where $\bm{1}_{a(t)=1}$ is the indicator function for the event that action at time $t$ is $a(t)=1$. In (\ref{eq:5}) we assume that the initial age of the system is $v$ which is independent of the policy, and initially the the transmitter has no packet to transmit.

\textit{Transition probability:} We denote the transition probability from state $s$ to state $s'$, under an action $a\in{\{0,1\}}$, with $P_{a}(s,s')$. Let us consider the following states,
\begin{align}
    & s_{1}=(v_{1}+1,1), \ s_{2} = (v_{1}+1,v_{2}+1), \nonumber\\ 
    & s_{3} = (1,\infty), \ s_{4} = (v_{2}+1,\infty).
\end{align}
For a state $s=(v_{1},v_{2})$, we have the following transition probabilities,
\begin{align}
    &P_{1}(s,s_{1}) = (1-q), \ P_{1}(s,s_{3}) =q, \nonumber\\
    & P_{0}(s,s_{2}) = (1-q), \ P_{0}(s,s_{4}) = (1-q),
\end{align}
and any other transition probability is $0$.

For the presentation of the next theorem we define the following classes of policies \cite{sennott1993constrained}. We say that a policy $f$ is a stationary policy if it is non-randomized, and its decision at time $t$ only depends on the state of the system and is independent of the time index $t$. We denote the action taken by the policy for a state $s\in{\mathcal{S}}$ as $f(s)$. Assume that $f$ and $g$ are two stationary policies, such that two policies differ only for one state $s' \in{\mathcal{S}}$, i.e., $f(s) = g(s) $, if $s \neq s'$. Let us consider $0<p<1$, such that at time $t=1$, we choose the policy $f$ and stick to it for the whole time horizon $T$ with probability $p$, or choose the policy $g$ and stick to it for the whole time horizon $T$ with probability $(1-p)$. Then, such a policy is said to be a Bernoulli-modulated simple policy. 

In Theorem~\ref{th:1}, we show that there exists a Bernoulli-modulated simple policy that is optimal for the problem in (\ref{eq:5}). 

\begin{theorem}\label{th:1}
    There exists a Bernoulli-modulated simple policy which is optimal for the following problem,
 \begin{align} 
    &\inf_{\pi\in{\Pi}}  \limsup_{T\rightarrow \infty} \frac{1}{T} \sum_{t=1}^{T} \mathbb{E}_{\pi} [C(s(t),a(t))|s(1)=(v,\infty)] \nonumber\\ & \textrm{s.t.}  \quad \liminf_{T\rightarrow \infty } \frac{1}{T} \sum_{t=1}^{T} \mathbb{E}_{\pi}[\bm{1}_{a(t)=1}|s(1)=(v,\infty)] \leq f_{\max}.\label{eq:prob1}
\end{align}
\end{theorem}

Therefore, from the Theorem~\ref{th:1}, we restrict the policy space $\Pi$ in (\ref{eq:5}), to the set of all Bernoulli-modulated simple policies, which we denote with $\Pi_{1}$.  Next, following the renewal theory \cite{ross1995stochastic, durrett2019probability}, we show that both of the limits in (\ref{eq:5}) exist. 

\begin{theorem}\label{th:3}
    Consider a Bernoulli-modulated simple policy $\pi$. Then, under the policy $\pi$, the following two limits exists,
    \begin{align}
         & \limsup_{T\rightarrow \infty} \frac{1}{T} \sum_{t=1}^{T} \mathbb{E}_{\pi} [C(s(t),a(t))|s(1)=(1,\infty)], \label{eq:13} \\ &   \liminf_{T\rightarrow \infty } \frac{1}{T} \sum_{t=1}^{T} \mathbb{E}_{\pi}[\bm{1}_{a(t)=1}|s(1)=(1,\infty)]. \label{eq:14}
    \end{align}
\end{theorem}

Now, we consider the dual MDP formulation of (\ref{eq:5}). As both of the limits in (\ref{eq:3}) exist, for a given Lagrangian coefficient $\lambda\geq 0$, we define the dual MDP of (\ref{eq:3}) as,
\begin{align}\label{eq:33}
    \inf_{\pi\in{\Pi_{1}}} \lim_{T\rightarrow \infty} \frac{1}{T} \sum_{t=1}^{T} &\mathbb{E}_{\pi}[C(s(t),a(t)) + \lambda \bm{1}_{a(t)=1}|s(1) = (1,\infty)]. 
\end{align}
We formulate (\ref{eq:33}) as an unconstrained MDP, with the same state space, action space, and transition probabilities, as the constrained MDP of (\ref{eq:3}). The cost of this new MDP is defined as follows,
\begin{align}
    C_{1}(s(t),a(t);\lambda) = C(s(t),a(t)) + \lambda \bm{1}_{a(t)=1}.
\end{align}
Thus, the dual MDP in (\ref{eq:33}) can be reformulated as,
\begin{align}\label{eq:jr25}
     \inf_{\pi\in{\Pi_{1}}} \lim_{T\rightarrow \infty} \frac{1}{T} \sum_{t=1}^{T} &\mathbb{E}_{\pi}[C_{1}(s(t),a(t);\lambda) |s(1) = (1,\infty)]. 
\end{align} 
We have $\lambda$ in the cost expression of the dual MDP to express the fact that the cost of the dual MDP is dependent on $\lambda$. Next, we find several structural results of the dual MDP, which hold for any arbitrary $\lambda>0$. Thus, for the simplicity of presentation, whenever it is not necessary to explicitly mention $\lambda$, we omit it from the cost expression.

For the compactness of this paper, next, we introduce the discounted cost MDP and discounted value function corresponding to the problem of this paper, and study some of the properties of the discounted value function, which is available in the literature, see \cite{bertsekas2012dynamic}, \cite{sennott1989average}.  We extensively use those properties to prove some of the fundamental results of an optimal sampling policy. For, $0<\alpha<1$, $s'\in{\mathcal{S}}$ and for an arbitrary policy $\pi$ we define $V_{\alpha}^{\pi}(s)$ as,
\begin{align}\label{eq:icc14}
    V_{\alpha}^{\pi}(s') =  \sum_{t=1}^{\infty} \alpha^{t} \mathbb{E}_{\pi} [C_{1}(s(t),a(t))|s(1)=s'].
\end{align}
We denote $V_{\alpha}(s')$ with,
\begin{align}
    V_{\alpha}(s') = \inf_{\pi} V_{\alpha}^{\pi}(s').
\end{align}
From \cite{sennott1989average}, we say that if $V_{\alpha}(s')$ is finite for all $s'\in{\mathcal{S}}$, then the following relation holds,
\begin{align}
    V_{\alpha}(s') = \min_{a\in{\{0,1\}}} V_{\alpha}(s';a),
\end{align}
where
\begin{align}
    V_{\alpha}(s';a) = C_{1}(s',a) + \alpha\sum_{s''\in{\mathcal{S}}}P_{a}(s',s'') V_{\alpha}(s'').  
\end{align}
Finally, let us define $V_{\alpha,0}(s')=0$, for all $s'\in{\mathcal{S}}$, and consider the following iteration,
\begin{align}\label{eq:31}
    V_{\alpha,n+1}(s') = \min_{a\in{\{0,1\}}}\big\{ C_{1}(s',a) + \sum_{s''\in{\mathcal{S}}} P_{a}(s',s'')V_{\alpha,n}(s'')\big\}.
\end{align}
From \cite{sennott1989average}, we say that,
\begin{align}\label{eq:32}
    \lim_{n\rightarrow\infty} V_{\alpha,n}(s) = V_{\alpha}(s).
\end{align}

In Theorem~\ref{th:dualstat}, we show that there exists a deterministic policy that solves (\ref{eq:33}).

\begin{theorem}\label{th:dualstat}
    For  $\lambda\geq 0$, there exists a stationary policy that optimally solves the problem in (\ref{eq:jr25}).
\end{theorem}

\begin{remark}
    We use Theorem~\ref{th:dualstat}, to connect the discounted cost problem to the average cost problem. Whichever property we prove for an optimal policy for the discounted cost problem in (\ref{eq:icc14}) is also applicable to the average cost problem in (\ref{eq:jr25}). 
\end{remark}

In the next lemma, we examine the monotonicity property of the value function $V_{\alpha}(s)$, with respect to $v_{1}$ and $v_{2}$.

\begin{lemma}\label{lemma:1}
    For $0<\alpha<1$ and $s=(v_{1},v_{2})$, $V_{\alpha}(s)$ is a monotonically increasing function of $v_{1}$ and $v_{2}$. Similarly, for a state $s'=(v_{1},\infty)$, $V_{\alpha}(s')$ is an increasing function of $v_{1}$.
\end{lemma}

Now, we use the techniques of \cite{banerjee2023re,banerjee2024preempt,banerjee2024wiopt} to extend the countable state space $\mathcal{S}$, to an uncountable state space, which will be useful to prove future results. Specifically, we assume that the age of the monitor $v_{1}$ and the age of the packet at the transmitter $v_{2}$ can take values from the set of real numbers. We denote this new state space with $\bar{\mathcal{S}}$. Thus,
\begin{align}
    \bar{\mathcal{S}} = \big\{&(v_{1},v_{2})\cup (v_{1},\infty), (v_{1},v_{2})\in{\mathbb{R}^{2}}, 1\leq v_{2}\leq v_{1}\big\}.
\end{align}
With the state space $\bar{\mathcal{S}}$, the same action space, cost, and transition probability of the MDP $\Delta$, We denote a new MDP with $\bar{\Delta}$. We denote the $\alpha$ optimal value function on $\bar{\mathcal{S}}$ with $\bar{V}_{\alpha}(s)$, $s\in{\bar{\mathcal{S}}}$. Note that, due to the structure of the transition probability of $\bar{\Delta}$, the evolution of states under a stationary policy always belongs to a countable subset of $\bar{\mathcal{S}}$. This sub-set changes with the initial state accordingly. For example, consider a state $s=(4.5, 3.5)\in{\bar{\mathcal{S}}}$. Let us consider the following sets,
\begin{align}
    & \mathcal{S}_{1} = \{(4.5+x,3.5+x), x\in{\mathbb{N}}\}, \\ &\mathcal{S}_{2} = \{(3.5+x,\infty), x\in{\mathbb{N}}\},\\ &\mathcal{S}_{3} = \{(4.5+x,y), (x,y)\in{\mathbb{N}^{2}}\}.
\end{align}
Note that under any policy the evolution of state $s$ belongs to the set $\mathcal{S}\cup\mathcal{S}_{1}\cup\mathcal{S}_{2}\cup\mathcal{S}_{3}$, which is a countable subset of $\bar{\mathcal{S}}$.   Thus, for any $s\in{\bar{\mathcal{S}}}$, and $0<\alpha<1$, if $\bar{V}_{\alpha,n}(s)$ follows the similar iteration of (\ref{eq:31}). Then,
\begin{align}\label{eq:new134}
    \lim_{n\rightarrow\infty} \bar{V}_{\alpha,n}(s) = \bar{V}_{\alpha}(s).
\end{align}
Also note that for any $s\in{\mathcal{S}}$, the following relation holds,
\begin{align}\label{eq:new135}
    \bar{V}_{\alpha}(s) = {V}_{\alpha}(s).
\end{align}
 
\begin{lemma}\label{lemma:2}
    For $0<\alpha<1$ and $s\in{\bar{\mathcal{S}}}$,  $\bar{V}_{\alpha}(s)$ is a coordinate wise concave function with respect to $v_{1}$ as well as of $v_{2}$.
\end{lemma}

\begin{remark}
    As, $\bar{V}_{\alpha}(s)$ is a concave function of $v_{1}$ and $v_{2}$, the left-sided derivative of $\bar{V}_{\alpha}(s)$, with respect to $v_{1}$ and $v_{2}$ exists, where we define the left-sided derivative as follows,
\begin{align}
     \!\!\pdv{\bar{V}_{\alpha,n}(s)^{-}} {v_{1}}\!\! =\!\! \lim_{h\rightarrow 0+} \frac{\bar{V}_{\alpha,n}((v_{1},v_{2})) - \bar{V}_{\alpha,n}((v_{1}-h,v_{2}))}{h}.
\end{align}
\end{remark}

\begin{theorem}\label{th:5}
    For $0<\alpha<1$, $\lambda\geq 0$, and for $s=(v_{1},v_{2})\in{\bar{\mathcal{S}}}$, $s'=(v_{1}',v_{2}')\in{\bar{\mathcal{S}}}$, where $v_{2}, v_{2}' \neq \infty$, the following relation holds for all $n\in{\mathbb{N}}$, 
    \begin{align}\label{eq:41}
        \pdv{\bar{V}_{\alpha,n}(s)^{-}} {v_{1}} =  \pdv{\bar{V}_{\alpha,n}(s')^{-}} {v_{1}}.
    \end{align}
\end{theorem}

\begin{corollary}\label{corr:1}
    For $0<\alpha<1$, $\lambda\geq 0$, and for $s=(v_{1},v_{2})\in{\bar{\mathcal{S}}}$, $s'=(v_{1}',v_{2}')\in{\bar{\mathcal{S}}}$, where $v_{2}, v_{2}' \neq \infty$, the following relation holds,
    \begin{align}
         \pdv{\bar{V}_{\alpha}(s)^{-}} {v_{1}} =  \pdv{\bar{V}_{\alpha}(s')^{-}} {v_{1}}.
    \end{align}
\end{corollary}

\begin{remark}
    Theorem~\ref{th:5} and Corollary~\ref{corr:1}, show that the slope of the value function is independent of the state, if the transmitter has a packet to transmit, i.e., $v_{2}\neq \infty$.
\end{remark}

Next, we show that corresponding to the discounted value function, the transmitter having no packet is equivalent to the transmitter having a packet of the age of the monitor and keeps transmitting it. 

\begin{lemma}\label{lemma:3}
    For $0<\alpha<1$, $v_{1}\in{\mathbb{N}}$, and $n\in{\mathbb{N}}$, the following holds, 
    \begin{align}\label{eq:47}
        V_{\alpha,n}((v_{1},\infty)) = V_{\alpha,n}((v_{1},v_{1})).
    \end{align}
\end{lemma}

Next, we show that when $\lambda=0$, action $a=1$ is optimal for any state $s\in{\mathcal{S}}$, corresponding to the discounted value function.

\begin{lemma}\label{lemma:4}
    For $0<\alpha<1$ and $\lambda=0$, the following relation holds for all $s=(v_{1},v_{2})\in{{\mathcal{S}}}$,
    \begin{align}\label{eq:53}
        V_{\alpha}(s;1) \leq V_{\alpha}(s;0).
    \end{align}
\end{lemma}

\begin{corollary}\label{corr:2}
    For $0<\alpha<1$, and for $s=(v_{1},v_{2})\in{\bar{\mathcal{S}}}$, where $v_{2}\neq \infty$, the following relation holds,
    \begin{align}
        \pdv{\bar{V}_{\alpha,n}(s;0)^{-}} {v_{1}} &= \pdv{\bar{V}_{\alpha,n}(s;1)^{-}} {v_{1}},\label{eq:67} \\ \pdv{\bar{V}_{\alpha}(s;0)^{-}} {v_{1}} &= \pdv{\bar{V}_{\alpha}(s;1)^{-}} {v_{1}}.\label{eq:68}
    \end{align}
\end{corollary}

\begin{remark}
    We prove Corollary~\ref{corr:2}, by using Theorem~\ref{th:5} and Corollary~\ref{corr:1}. Corollary~\ref{corr:2} implies that the value functions corresponding to actions $a=0$ and $a=1$ cannot cross each other if the transmitter has a packet to transmit.
\end{remark}

\begin{corollary}\label{corr:n3}
    If $a=1$ is optimal for a state $s=(v_{1},v_{2})$, then $a=1$ is also optimal for state $s'=(v_{1}',v_{2})$, where $s,s'\in{\bar{\mathcal{S}}}$. We consider the optimality with respect to the discounted value function. 
\end{corollary}

\begin{remark}
    Corollary~\ref{corr:n3} is a direct consequence of Corollary~\ref{corr:2}.
\end{remark}

\begin{lemma}\label{lemma:5}
    For $0<\alpha<1$, $s=(v_{1},v_{2})$, and $s'=(v_{1},\infty)$, where $s,s'\in{\bar{\mathcal{S}}}$, the following relation holds for $n\in{\mathbb{N}}$,
    \begin{align}\label{eq:new73}
          \pdv{\bar{V}_{\alpha,n}(s)^{-}} {v_{1}} \leq   \pdv{\bar{V}_{\alpha,n}(s')^{-}} {v_{1}}.
    \end{align}
\end{lemma}

\begin{corollary}\label{corr:3}
    If $a=1$ is optimal for a state $s=(v_{1},\infty)$, then, $a=1$ is also optimal for a state $s'=(v_{1}+x,\infty)$, for $x>0$. Here, we consider the optimality with respect to the discounted value function.
\end{corollary}

\begin{remark}
    Corollary~\ref{corr:3}, provides a threshold structure on the age of the monitor when the transmitter has no packet to transmit. We prove Corollary~\ref{corr:3} with Lemma~\ref{lemma:5}.
\end{remark}

In the next lemma, we find an upper bound on $\pdv{\bar{V}_{\alpha}(s)^{-}} {v_{1}}$, which will be useful to obtain an optimal policy.

\begin{lemma}\label{lemma:6}
    For $0<\alpha<1$, and $s\in{\bar{\mathcal{S}}}$,
    \begin{align}\label{eq:82}
        \pdv{\bar{V}_{\alpha}(s)^{-}} {v_{1}} \leq  \frac{1}{1-\alpha} + \frac{1}{1-\alpha(1-q)}.
    \end{align}
\end{lemma}

Like Lemma~\ref{lemma:2}, we present the concavity of the discounted value function with respect to $\lambda$, the proof of which is similar to the proof for Lemma~\ref{lemma:2}.

\begin{lemma}\label{lemma:7}
    For $0<\alpha<1$, $\lambda\geq 0$, and $s\in{\bar{\mathcal{S}}}$, $\bar{V}_{\alpha}(s,\lambda)$ is a coordinatewise concave function with respect to $\lambda$.
\end{lemma}

Like Lemma~\ref{lemma:1}, we present the monotonicity of the discounted value function of $\lambda$.

\begin{lemma}
    For $0<\alpha<1$, $\lambda\geq 0$, and $s\in{\bar{\mathcal{S}}}$, $\bar{V}_{\alpha}(s,\lambda)$, is an increasing function of $\lambda$.
\end{lemma}

\begin{theorem}\label{th:6}
    For a state $s=(v_{1},\infty)\in{\bar{\mathcal{S}}}$, there exists a $\lambda\geq 0$, such that,
    \begin{align}\label{eq:91}
        \bar{V}_{\alpha}(s,\lambda;0)=\bar{V}_{\alpha}(s,\lambda;1).
    \end{align}
\end{theorem}

\begin{remark}
    Theorem~\ref{th:6} says that when the transmitter has no packet to transmit, then for any arbitrary age of the monitor, there exists a $\lambda\geq 0$, such that both actions $a=0$ and $a=1$ are equally likely to get chosen by an optimal sampling policy. 
\end{remark}

\begin{theorem}\label{th:7}
    For $0<\alpha<1$, $n\in{\mathbb{N}}$, $s=(v_{1},\infty)$, and $s'=(v_{1},v_{2})$, 
 $\pdv{\bar{V}_{\alpha,n}(s)^{-}} {v_{1}}$ and $\pdv{\bar{V}_{\alpha,n}(s)^{-}} {v_{1}}$ are increasing functions of $\alpha$. Similarly, $\pdv{\bar{V}_{\alpha,n}(s)^{-}} {v_{1}}-\pdv{\bar{V}_{\alpha,n}(s')^{-}} {v_{1}}$ is also an increasing function of $\alpha$.
\end{theorem}

\begin{lemma}\label{lemma:9}
    For $0<\alpha<1$, $s=(v_{1},\infty)\in{\bar{\mathcal{S}}}$, $\pdv{\bar{V}_{\alpha}(s)^{-}} {v_{1}}\geq 1$.
\end{lemma}

\begin{lemma}\label{lemma:10}
    For a fixed $\lambda\geq 0$, $0<\alpha<1$, and $s=(v_{1},\infty)\in{\bar{\mathcal{S}}}$, $\bar{V}_{\alpha}(s;0) - \bar{V}_{\alpha}(s;1)$ is a strictly increasing function of $v_{1}$.
\end{lemma}

\begin{theorem}\label{th:8}
For $s=(v_{1},\infty)\in{\bar{\mathcal{S}}}$ and $0<\alpha<1$, let us denote $\lambda_{\alpha}(s)$ as,
\begin{align}
    \lambda_{\alpha}(s) = \inf\{\lambda|\bar{V}_{\alpha}(s,\lambda;0) = \bar{V}_{\alpha}(s,\lambda;1)\}.
\end{align}Then,
     $\lambda_{\alpha}(s)$ is an increasing function of $\alpha$. Moreover, $\lambda_{\alpha}(s)$ is finite.
\end{theorem}

From Theorem~\ref{th:8}, we say that, $\lim_{\alpha\rightarrow 1} \lambda_{\alpha}(s)$, exists and finite, we denote it with $\bar{\lambda}(s)$. For a given $\alpha$, and for any $s=(v_{1},\infty)\in{\bar{\mathcal{S}}}$, if we choose $\lambda$ to be $\lambda_{\alpha}(s)$. Then, according to Theorem~\ref{th:8}, Lemma~\ref{lemma:10} and Corollary~\ref{corr:3}, we say that when the transmitter has no packet to transmit, it is optimal for the sampler not to sample a packet till the age of the monitor reaches the age $v_{1}$. When the monitor age is $v_{1}$, both actions $a=0$ and $a=1$ are optimal for the sampler. When the age of the monitor is more than $v_{1}$, action $a=1$ becomes optimal. 

Consider a sequence $\{\alpha_{n}\}_{n=1}^{\infty}$, such that $\lim_{n\rightarrow \infty} \alpha_{n} = 1$. Note that, for a given $\alpha$, and $\lambda$, a stationary deterministic policy $\pi$ can be expressed as a countable infinite dimension matrix, where the entry on the $v_{1}$th row and the $v_{2}$th column represent the action of policy $\pi$ for the state $(v_{1},v_{2})$. Now, we denote an optimal policy corresponding to the discounted cost for $\alpha_{n}$ and $\lambda_{\alpha_{n}}(s)$ with $\pi_{n,s}$. Now, we know that a state $(v_{1},v_{2})$, where $v_{2}>v_{1}$, is not feasible. Thus, for the $i$th row of the matrix $\pi_{n,s}$, we can assign any value for entries $(i,j)$, where $j>i$; without loss of generality we assign $0$ to those entries. Let us consider that $s=(v_{1},\infty)$. Then, we know that for any state $s'=(v_{1}',v_{2}')$, action $a=0$ is optimal, if $v_{2}'<v_{1}$, action $a=1$ optimal if $v_{1}<v_{2}'$. Both actions $a=0$ and $a=1$ are optimal, if $v_{1}=v_{2}'$, and without loss of generality, we assign $a=1$ to those entries. Thus, for all $n$, $\bar{\pi}_{n,s}$ are identical. Thus, we remove the dependence with $n$, and denote it with $\bar{\pi}_{s}$. Thus, $\bar{\pi}_{s}$ is optimal corresponding to the discount cost, for $\alpha$ being any element of the sequence $\{\alpha_{n}\}_{n=1}^{\infty}$ and $\lambda$ being $\lambda_{\alpha_{n}}(s)$. 

Now, we show that $\bar{\pi}_{s}$  is optimal for (\ref{eq:jr25}), when $\lambda=\bar{\lambda}(s)$.  

\begin{theorem}\label{th:9}
    The policy $\bar{\pi}_{s}$ is optimal for
    \begin{align}
         \inf_{\pi\in{\Pi_{1}}} \lim_{T\rightarrow \infty} \frac{1}{T} \sum_{t=1}^{T} &\mathbb{E}_{\pi}[C_{1}(s(t),a(t);\bar{\lambda})|s(1) = (1,\infty)]. 
    \end{align}    
\end{theorem}

Let us assume that $s=(v_{1},\infty)\in{\mathcal{S}}$, then under the policy $\bar{\pi}_{s}$, the sampler only samples a packet, if either there is a packet at the transmitter and the age of that packet is greater than or equal to $v_{1}$, or there is no packet at the transmitter and the age of the monitor is greater than or equal to $v_{1}$. As we assume that the initial state is $(1,\infty)$, under the policy $\bar{\pi}_{s}$, the sampler samples a fresh packet in every $v_{1}$ interval. Thus, 
\begin{align}
    \lim_{T\rightarrow \infty } \frac{1}{T} \sum_{t=1}^{T} \mathbb{E}_{\pi}[\bm{1}_{a(t)=1}|s(1)=(1,\infty)] =\frac{1}{v_{1}}.
\end{align}

Now, we define a policy $\pi_{R}$ as follows: If we can find an integer $v_{1}$, such that $\frac{1}{v_{1}} = f_{\max}$, then $\pi_{R}=\bar{\pi}_{(v_{1},\infty)}$. Otherwise, we can find an integer $v_{1}$, such that, $\frac{1}{v_{1}+1} \leq f_{\max}\leq\frac{1}{v_{1}}$. Then, under the policy $\pi_{R}$, at time $t=1$, the sampler chooses the policy $\bar{\pi}_{(v_{1},\infty)}$ with probability $p$ and chooses the policy $\bar{\pi}_{(v_{1}+1,\infty)}$ with probability $(1-p)$, and stick to the chosen policy for the whole time horizon $T$, where $p$ satisfies,
\begin{align}\label{eq:new127}
     \frac{p}{v_{1}} +  \frac{(1-p)}{v_{1}+1} = f_{\max}.
\end{align}
\begin{theorem}
    The policy $\pi_{R}$ is optimal for the problem,
\begin{align} 
    &\inf_{\pi\in{\Pi_{1}}}  \lim_{T\rightarrow \infty} \frac{1}{T} \sum_{t=1}^{T} \mathbb{E}_{\pi} [C(s(t),a(t))|s(1)=(v,\infty)] \nonumber\\ & \textrm{s.t.}  \quad \lim_{T\rightarrow \infty } \frac{1}{T} \sum_{t=1}^{T} \mathbb{E}_{\pi}[\bm{1}_{a(t)=1}|s(1)=(v,\infty)] \leq f_{\max}.
\end{align}
\end{theorem}
 
\section{Numerical Results}
In this section, we compare the average age for various sampling policies. First, we introduce the relative value iteration-based method to find an optimal sampling policy. For a specific $\lambda$, let us consider the following iteration on $n\in{\mathbb{N}}$, and for $s\in{\mathcal{S}}$,
\begin{align}
    &h^{n+1}(s) \!=\! (1\!-\!\gamma) h^{n}(s) \!+\! \min_{a} \{C_{1}(s,a;\lambda)\!+\! \gamma\!\sum_{s'\in{\mathcal{S}}}\! P_{a}(s,s')\nonumber\\& h^{n}(s')\}  - \min_{a}\{C_{1}(t,a;\lambda) + \gamma \sum_{s'\in{\mathcal{S}}} P_{a}(t,s') h^{n}(s')\},
\end{align}
where $0<\gamma<1$, $t=(1,\infty)$, and $h_{0}(s)=0$  for all $s\in{\mathcal{S}}$. From \cite{bertsekas2012dynamic}, we say that,
\begin{align}\label{eq:n161}
    &\lim_{n\rightarrow\infty} h^{n}(s) =  \frac{h^{*}(s)}{\gamma}, \\ &\lim_{n\rightarrow\infty} \min_{a}\{C_{1}(t,a;\lambda) + \gamma \sum_{s'\in{\mathcal{S}}} P_{a}(t,s') h^{n}(s')\}=\tau,
\end{align}
where $\tau$ is the optimal average cost for (\ref{eq:jr25}), and and $h^{*}(s)$ satisfies the following relation,
\begin{align}\label{eq:195}
    h^{*}(s) +\lambda = \min_{a} \{C(s,a)+\sum_{s'\in{\mathcal{S}}} P_{a}(s,s') h^{*}(s')\}.
\end{align}

\begin{figure}[t]
    \centerline{\includegraphics[width = 1.1\columnwidth]{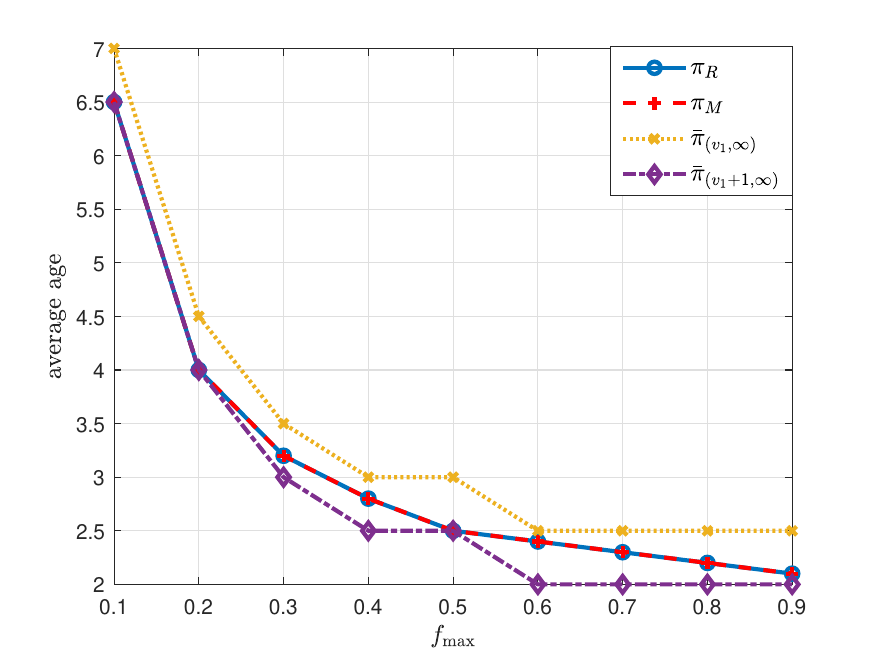}}
    \caption{We compare the average age for different policies for $q=0.5$.}
    \label{fig:3}
\end{figure}

A stationary policy which satisfies the minimization of (\ref{eq:195}) is optimal for (\ref{eq:jr25}), which we call as $\pi^{*}_{\lambda}$. We denote the average sampling cost for $\pi^{*}_{\lambda}$ with $s_{\lambda}$. Now, we run a bi-section search on $\lambda$. We initialize with $\lambda^{+}$, being a large number and $\lambda^{-}=0$.  We denote $\lambda^{m} = \frac{\lambda^{+}+\lambda^{-}}{2}$. Now, we find $\pi^{*}_{\lambda^{m}}$ and $s_{\lambda^{m}}$, by running (\ref{eq:195}). If $s_{\lambda}^{m}< f_{\max}$, we choose $\lambda^{+}=\lambda^{m}$, if $s_{\lambda^{m}}> f_{\max}$, we choose $\lambda^{-}=\lambda^{m}$, and if $s_{\lambda^{m}}=f_{\max}$, we choose $\lambda^{-}=\lambda^{m}=\lambda^{+}$. If $|\lambda^{+}-\lambda^{-}|\geq \epsilon$, we run the iteration (\ref{eq:195}) again, and the same process goes on, where $\epsilon$ is a system parameter. When we get $\lambda^{+}$ and $\lambda^{-}$ such that $|\lambda^{+}-\lambda^{-}|< \epsilon$,  the sampler chooses $\pi^{*}_{\lambda^{-}}$ with probability $p$ and $\pi^{*}_{\lambda^{+}}$ with probability $(1-p)$, at time $t=1$, and sticks to the chosen policy for the entire time horizon, where $p$ satisfies the following equation,
\begin{align}
    p{s_{\lambda^{-}}} + (1-p) s_{\lambda^{+}} = f_{\max}.
\end{align}
We call this policy as $\pi_{M}$. We see from Fig.~\ref{fig:3}, that the average ages of our proposed policy $\pi_{R}$ and $\pi_{M}$, overlap.

\bibliographystyle{unsrt}
\bibliography{references}

\end{document}